\documentclass[traditabstract]{aa} 

\usepackage[dvips]{graphicx}          
\usepackage[comma,authoryear]{natbib} 
\bibpunct{(}{)}{,}{a}{}{,}            

\begin{document}

   \title{Detecting individual gravity modes in the Sun}


   \author{R.A. Garc\'\i a\inst{1,2}
               \and
           J. Ballot\inst{3}
           \and
           A. Eff-Darwich\inst{4,5}
           \and
           R. Garrido\inst{6}
           \and
          A. Jim\'enez\inst{5,4}
          \and
          S. Mathis\inst{1}
          \and
          S. Mathur\inst{7}
          \and
          A. Moya\inst{6}
          \and
          P.L. Pall\'e\inst{5,4}
          \and
          C. R\'egulo\inst{5,4}
          \and
          D. Salabert\inst{5,4}
          \and
          J.C. Su\'arez\inst{6}
          \and
          S. Turck-Chi\`eze\inst{1}
          }

   \institute{Laboratoire AIM, CEA/DSM-CNRS, Universit\'e Paris 7 Diderot, IRFU/SAp, Centre de Saclay, F-91191 Gif-sur-Yvette, France \\
              \email{rgarcia@cea.fr}
         \and
           GEPI, Observatoire de Paris, CNRS, Universit\'e Paris Diderot; 5 place Jules Janssen, 92190 Meudon, France
         \and
          Laboratoire d'Astrophysique de Toulouse-Tarbes, Universit\'e de Toulouse, CNRS, 14 av. Edouard Belin, F-31400 Toulouse, France 
          \and
         Universidad de La Laguna, E-38205 La Laguna, Tenerife, Spain 
         \and
         Instituto de Astrof\'\i sica de Canarias,  E-38200 La Laguna, Tenerife, Spain    
         \and
         Instituto de Astrof\'\i sica de Andaluc\'\i a CSIC, Cno. Bajo de Huetor, 50, Granada, Spain
         \and
         Indian Institute of Astrophysics, Koramangala, Bangalore 560034, India
                      }

  \abstract
   {Many questions are still open regarding the structure and the dynamics of the solar core. By constraining more this region in the solar evolution models, we can reduce the incertitudes on some physical processes and on momentum transport mechanisms. A first big step was made with the detection of the signature of the dipole-gravity modes in the Sun, giving a hint of a faster rotation rate inside the core. A deeper analysis of the GOLF/SoHO data unveils the presence of a pattern of peaks that could be interpreted as dipole gravity modes. In that case, those modes can be characterized, thus bringing better constraints on the rotation of the core as well as some structural parameters such as the density at these very deep layers of the Sun interior. }


   \maketitle

\section{Introduction}
Over the last 40 years, the study of the solar low-degree acoustic modes has improved our knowledge of the Sun interior \citep{TouApp1997,ThiBou2000,GarReg2001,2004A&A...413.1135S,2009MNRAS.396L.100B}. In particular, the accurate characterization of these modes has allowed us to constrain very well the radiative zone below the tachocline down to the core, for both, its structure properties \citep[e.g.][]{2009ApJ...699.1403B} or its dynamics \citep{ChaEls2001,GarCor2004,2008SoPh..251..119G,2004A&A...425..229M,2005A&A...440..653M}. However, the detection of gravity (g) modes in the Sun is needed to progress in our knowledge of the inner radiative zone below 0.25 $R_\odot$ where about 50$\%$ of the mass is concentrated. For example, the detection of a few g modes in the inversions of the rotational profile would greatly improve the inferences at 0.1~$R_\odot$ \citep{2008A&A...484..517M, 2009arXiv0902.4142M}. 

There is a general consensus that the observational limit for low-degree, low-order p modes is around 1 mHz
\citep{BerVar2000,2008AN....329..461B,2009ApJ...696..653S}. This is due to the raising convective background \citep[e.g.][]{2008A&A...490.1143L} and the decreasing p-mode amplitudes as the frequency decreases. In order to go towards lower frequencies, we need to improve the signal-to-noise ratio (SNR) by: reducing the convective noise \citep[e.g.][]{GarJef1999}; combining several instruments \citep[e.g.][]{2007MNRAS.379....2B,2008arXiv0810.1696S} or using new instruments like PICARD \citep{2006cosp...36..170T} or GOLF-NG \citep{2006AdSpR..38.1812T}.

The expected g-mode surface amplitudes  are also very small. However, \citet{2009A&A...494..191B} have shown that these modes could have surface amplitudes of several mm/s  reaching the detectable regime of the GOLF instrument \citep{2009arXiv0910.0848A}. Indeed, using this instrument as well as VIRGO -- also on board SoHO -- we have been able to detect signals in the g-mode region that are very unlikely to be due to solar or instrumental noise \citep{2009ApJS..184..288J}. Even more, some patterns raise above the general convective noise level with a high confidence level \citep{STCGar2004,2008AN....329..476G}. However, the most important step forward was obtained by applying the same methodology used in asteroseismology to look for the large separation in solar-like stars with a very small SNR \citep[e.g.][]{2009arXiv0907.0608G}. Thus, by computing the power spectrum of the power spectrum (PSPS), \citet{2007Sci...316.1591G} were able to unveil the asymptotic periodicity of dipole g modes even without seeing a clear pattern of modes in the power spectrum (as it was, for example, the case of HD175726 where \citet{2009arXiv0908.2244M} detected the large separation without observing individual modes). The detailed study of this asymptotic periodicity reveals a higher rotation rate in the core of about 3-5 times in average faster than the rest of the radiative region and a better agreement in comparison with solar models computed with old surface abundances instead of the new ones \citep{2008SoPh..251..135G}. In this work, we study in details the power spectrum density (PSD) of the GOLF instrument to identify the pattern of peaks that produce the detected periodicity in the PSPS. If we succeed, these peaks could be the first g modes that can be correctly identified. Even more, we could analyze their properties, as for instance, their rotational splittings.

\section{Observations and analysis}

We have used 4472 days of GOLF \citep{GabGre1995} time series calibrated into velocity, starting April 11, 1996 \citep{GarSTC2005,UlrGar2000}. We have worked with a single -- full resolution -- power spectrum even knowing that GOLF has been observing in two different configurations with a different sensitivity to the visible solar disk \citep{GarRoc1998,1999A&A...348..627H}. To reduce the problem of the discretization in frequency, we have computed a 5 times zero-padded power spectrum 
\citep[see the discussion concerning this problem in][]{GabBau2002}. Finally, because the peaks could be spread over several bins \citep{2007Sci...316.1591G}, we have smoothed the PSD (with a boxcar of 41 nHz) to increase the SNR  as it is usually do in asteroseismology \citep[e.g.][]{2008Sci...322..558M}. 
   
 \section{Results}

\begin{figure}[!htbp]
     \begin{center}
       \includegraphics[angle=90,width=0.5\textwidth]{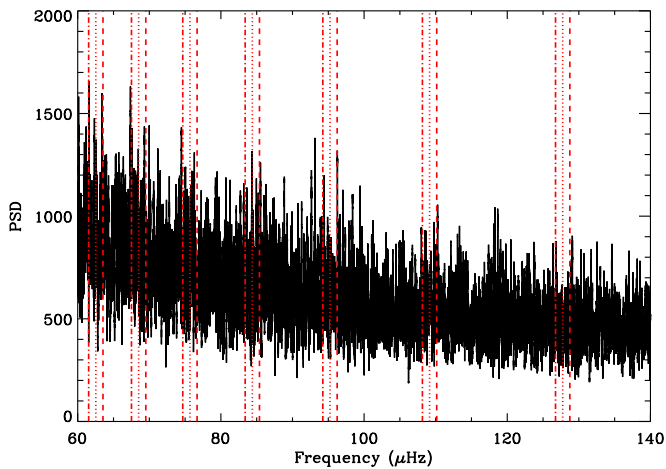}
     \end{center}
     \caption{GOLF PSD. The dotted vertical lines are the central frequencies of the dipole g modes computed by the Saclay seismic model. The vertical dashed and dot-dashed lines are the rotational split components with 4.5 times the rotational splitting of the radiative zone above 0.2 $R_{\odot}$.}
     \label{fig1}
   \end{figure}

Figure~\ref{fig1} shows the resulting PSD in the region between 60 and 140 $\mu$Hz. The vertical dotted lines are the central frequencies of the dipole g modes computed using the Saclay seismic model \citep{STCCou2001} with an updated grid of points. Moreover, changes in the solar surface abundances or in some of the ingredients of the solar model induce a small shift in the mode frequencies in this frequency range \citep{2007ApJ...668..594M,2007A&A...469.1145Z}.
  
We have used those computed frequencies as a guidance for the search. We have first verified that these theoretical predictions lie in the regions where the pattern generating the PSPS detected by  \citet{2007Sci...316.1591G} were located. Then, the rotational splittings were varied over the range given by \citet{2007Sci...316.1591G} (3 to 5 times the rotation rate in the radiative zone, $\Omega_{rad}$=433 nHz). Most of the split components match the highest peaks in the PSD for a rotation rate $\sim$4.5 $\Omega_{rad}$.
Work is still in progress, but if this is confirmed, these peaks could be the first undoubted g modes observed individually in the Sun.

\begin{acknowledgements}
Participation at the Ponte de Lima workshop was supported by the European Helio- and Asteroseismology Network (HELAS), a major international collaboration funded by the European Commission's Sixth Framework Programme.  SoHO  is a space mission of international cooperation between ESA and NASA. D. S. acknowledges the support of the grant PNAyA2007-62650 from the Spanish National Research Plan, R.A.G. the support of the French CNES/GOLF grant at the SAp and AM the support from a {\em Juan de la Cierva} contract of the Spanish Ministry of Science and Innovation.
\end{acknowledgements}

\bibliographystyle{aa}
\bibliography{/Users/rgarcia/Desktop/BIBLIO} 
\end{document}